\documentstyle[epsf]{mn}
\begin{document}

% ***************************TITLE ************************

\title[Observations of the Be/X-ray binary 4U1145-619]
    {Multiwavelength observations of the Be/X-ray binary 4U1145-619}

%****************************AUTHORS**********************

\author[J.B.Stevens  et al.]
{J.B.Stevens$^{1}$, P.Reig$^{1}$, M.J.Coe$^{1}$, D.A.H.Buckley$^{2}$,
J.Fabregat$^{3}$, I.A.Steele$^{4}$\\
$^{1}$Physics and Astronomy Department, The University, Southampton SO17 1BJ\\
$^{2}$South African Astronomical Observatory, P.O. Box 9, Observatory,
7935, South Africa\\
$^{3}$Departmento de Astronomia, Universidad de Valencia, 46100
Burjassot, Spain\\
$^{4}$Astrophysics Group, Liverpool John Moores University,
Liverpool, L3 3AP}

\date{Accepted \\
Received : Version \today \\
In original form ..}
\maketitle

% *************************** ABSTRACT *******************

\begin{abstract} 
We report optical and infrared observations of the massive X-ray
binary system 4U1145-619 (V801 Cen) which show that the circumstellar disc of the
Be star component is in decline. Infrared {\it JHKL}
magnitudes of V801Cen have been monitored from 1993 March to 1996
April. H$\alpha$ spectra have been obtained
throughout the same period. We find that both
the infrared excess and the Balmer emission have been in decline
throughout the period of observations. A 13 year optical and X-ray
history of the source has been collated, revealing a possible
correlation between the optical and X-ray activity. In
addition, we have used {\it uvby$\beta$} indices, corrected for both
circumstellar and interstellar effects, to calculate the physical
parameters of the underlying B star.

\end{abstract}
\newpage

%****************************KEYWORDS********************


% *************************** INTRODUCTION **************

%**************************** Be/XRAY BACKROUND**********

\section{Introduction}
The Be/X-ray binary systems represent the largest subclass of High
Mass X-ray Binaries (HMXB's). These systems consist of a compact object
(usually a neutron star) in a wide eccentric orbit with a Be star.
A Be star is defined to be an early type luminosity class III-V star,
which has at
some time shown emission in the Balmer lines \cite{jsj} . The Balmer emission, along with a significant
infrared excess is believed to originate in the circumstellar
material surrounding the Be star, probably in the form of an
equatorial disc. The X-ray emission in these systems is the result
of accretion of material onto the neutron star from the Be star's envelope. In the Be/X-ray
systems, mass transfer is enhanced during periastron passage, as the
neutron star passes through the denser regions of the companion's
disc, resulting in X-ray outbursts with luminosities typically 10-100
times stronger than the quiescent level \cite{bmc} . In some cases outbursts are seen with no correlation to orbital
phase (GRO J1948-03, Zhang et al., 1996; V0332+53, Terrel \&
Preidhorsky 1984). These outbursts are usually much more intense and
are believed to be the
result of large mass loss events from the Be star, although the
mechanism behind such events is yet unclear.\\
\hspace*{0.5cm}

%***************************SOURCE HISTORY*******************

The Be/X-ray binary 4U1145-619 is a highly variable X-ray source which
has been optically identified with the 9th magnitude B1{\sc V}e star
V801Cen (Bradt et al. 1977; Dower et al. 1978). White et al. (1978) found two X-ray pulsation periods of 292s and 297s from
the field of the {\it Uhuru} source 4U1145-61, but were unable to
confirm whether both periods originated in the one source, or whether
there were in fact two X-ray pulsars within the field. The ambiguity
was resolved when observations with the imaging proportional counter
on the {\it Einstein Observatory} (Lamb et al. 1980; White et al. 1980) revealed the presence of two pulsars separated by
15$^{\prime}$. The source identified with V801Cen (redesignated
4U1145-619) was found to pulsate with the 292s period, whilst the 297s
period was found to originate in a new source, designated 1E
1145.1-6141, which was later found to be a member of the class of
HMXB's with supergiant companions (Hutchings, Crampton \& Cowley,
1981). Analysis of the long-term X-ray behaviour of 4U1145-619
revealed recurrent outbursts with a period of 186.5 days. Outbursts
are typically of 10 day duration with flux levels increasing by a
factor of $\sim$5 (Watson et al. 1981; Priedhosky \& Terrell 1983;
Warwick, Watson \& Willingdale 1985). During the majority of the observed outbursts,
the pulsed fraction of the total flux is relatively low. This, and the low luminosity of
the source led White et al. (1983) to list 4U1145-619 as one of the
best approximations to a spherically accreting binary system. The accepted model for this system is that of a
long period eccentric binary. The 186.5 day period in X-ray behaviour
is then the consequence of phase dependant accretion from the Be
star's circumstellar disc.

% ************************** OBSERVATIONS *********************

\section{Observations}

In this section we present our own infrared and optical observations
made between 1993 March and 1996 April. These observations were made
as part of the Southampton/Valencia/SAAO long term monitoring campaign of
High Mass X-ray Binaries (Coe et al. 1993; Reig et al. 1996b). In addition, we
present optical photometric data taken from the catalogue of ESO's
Long Term Photometry of Variables (LTPV) project (Sterken et al. 1995 and
references therein)
covering the period 1982--1994, and a summary of previously published
X-ray observations.

%*****************************INFRARED PHOTOMETRY********

\subsection{Infrared photometry}

%***************************** TABLE 2. IR PHOTOMETRY****

\begin{table*}
\centering
\caption{IR photometry of V801 Cen}
\begin{tabular}{l c c c c c c }

\hline
Date	& MJD	& J& H & K  & L & J-K\\
\hline
5 March 93   & 49052    & 8.24$\pm$0.03 & 7.93$\pm$0.03 &
7.53$\pm$0.02 & 6.98$\pm$0.05 & 0.71$\pm$0.04\\
8 March 93    &49055 & 8.25$\pm$0.02 & 7.95$\pm$0.02 & 7.54$\pm$0.02 &
6.95$\pm$0.03& 0.71$\pm$0.03\\
26 June 93     & 49165 & 8.42$\pm$0.02 & 8.10$\pm$0.02 & 7.67$\pm$0.02
& 7.07$\pm$0.05& 0.75$\pm$0.03\\
5 March 94      &49417 & 8.66$\pm$0.02 & 8.43$\pm$0.02 & 8.05$\pm$0.03&&0.61$\pm$0.04\\
30 June 94 &49534	&8.482$\pm$0.011 &8.268$\pm$0.011
&7.955$\pm$0.007 &7.464$\pm$0.017& 0.527$\pm$0.013\\
21 February 95&  49770      & 8.66$\pm$0.01 & 8.44$\pm$0.02 &
8.19$\pm$0.02 && 0.47$\pm$0.02\\
23 February 95 & 49772      & 8.67$\pm$0.01 & 8.45$\pm$0.01 &
8.19$\pm$0.01 & 7.70$\pm$0.06& 0.48$\pm$0.01\\
17 August 95 &49947 	& 8.95$\pm$0.01 & 8.81$\pm$0.01 & 8.63$\pm$0.01 &&0.32$\pm$0.01\\
6 April 96 &50180 & 8.682$\pm$0.028 & 8.562$\pm$0.009 & 8.395$\pm$0.011 &
8.279$\pm$0.058&0.287$\pm$0.030\\
\hline
\end{tabular}
\end{table*}

%***************************FIGURE 2. IR LIGHTCURVES*******

\begin{figure*}
\begin{center}
{\epsfxsize 0.99\hsize
\leavevmode
\epsffile{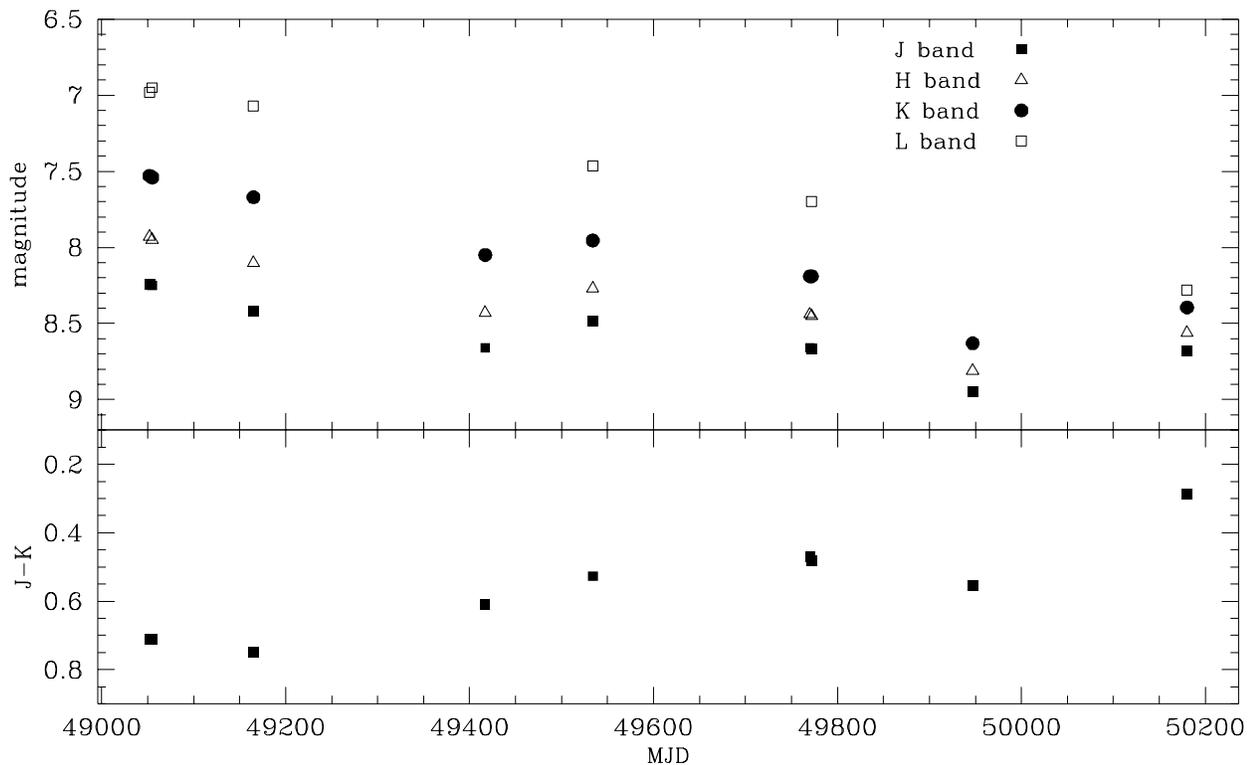}
}\end{center}
%\vspace{11cm}
\caption{IR lightcurves of 4U1145-619 during the period
1993-1996. A complete log of observations and associated errors is
found in Table 2.}
\label{ir}
\end{figure*}

%***************************GENERAL IR BLURB**************

Infrared data were obtained on ten occasions between 1993 March and
1996 April, using the MkIII IR photometer on the
1.9-m Cassegrain Telescope at the SAAO. The chopping secondary mirror
defines two effective apertures: on-source (star) and off-source
(background). Square apertures of 9 or 12 arcseconds were used, depending on the seeing. The source was placed
alternately in the two apertures for a number of cycles until a
sufficient precision was reached.  Observations were made in the SAAO {\it JHKL} system (1.25,
1.65, 2.2 and 3.5 $\mu$m; Carter 1990). A log of the observations, results and
corresponding errors is presented in Table 1, and lightcurves are
plotted in Figure~1.  The results show a clear
decline in all wavebands, with {\it J}
decreasing by $\sim$0.4-mag, and {\it L} decreasing by
$\sim$1.3-mag. In the same period of time, the {\it J-K}
index has decreased by $\sim$0.5. Further discussion of the IR data is
left until Section 4.

%**************************OPTICAL SPECTROSCOPY****************

%**************************FIGURE 1. OPTICAL SPECTRA***********

\begin{figure}
%\vspace{11cm}
\begin{center}{
\epsfxsize 1.0\hsize
\leavevmode
\epsffile{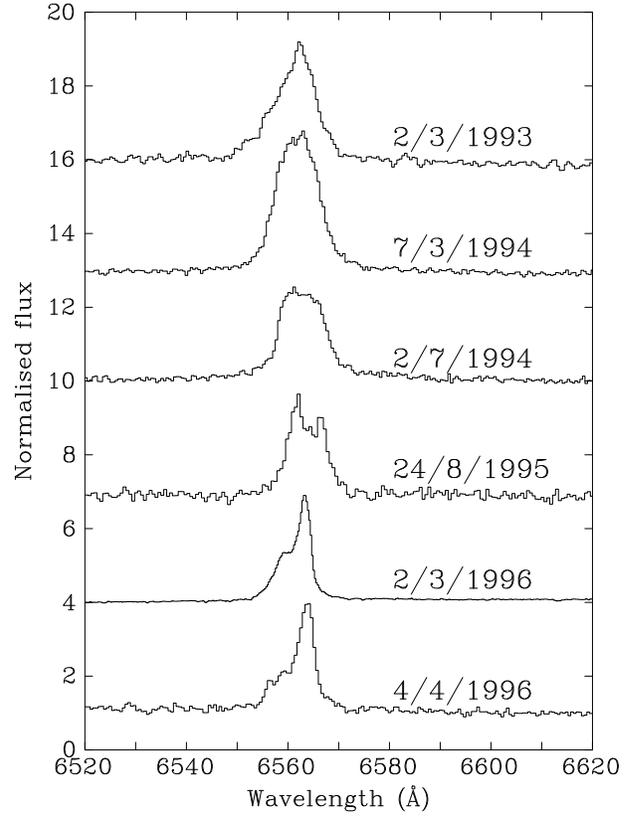}
}\end{center}
\caption{H$\alpha$ profiles of V801 Cen, the optical counterpart to
4U1145-619. The March 1996 spectrum was obtained at the AAT, all other
spectra are from the SAAO 1.9-m telescope. Fluxes are normalised
to the continuum and offset from each other to allow comparison.}
\label{alpha}
\end{figure}

%**************************TABLE 1. OPTICAL OBS***************

\begin{table}
\centering
\label{spec}
\caption{Optical spectroscopy of V801Cen. Errors in equivalent width
measurements are typically 10 per cent}
\begin{tabular}{l c c c}
\hline
Date &Site & \hspace{1.0cm}EW (\AA)\\
 & &H$\alpha$& H$\beta$\\
\hline
1993 Mar 2  & SAAO& -38& \\
1993 Mar 4  & SAAO& & -4.5\\
1994 Mar 7  & SAAO& -45& \\
1994 Jul 2  & SAAO& -25& \\
1994 Jul 3  & SAAO& -24  & \\
1994 Jul 4  & SAAO&  & -2.5\\
1995 Aug 24 & SAAO& -27& \\
1996 Mar 2  & AAT & -16& \\
1996 Apr 3  & SAAO & -13 &\\
1996 Apr 4  & SAAO & -14 &\\
1996 Apr 5  & SAAO &  & -1.5\\
\hline
\end{tabular}
\end{table}

\subsection{Optical spectroscopy}

We observed the source ten times with the 1.9-m telescope at the SAAO between 1993 March and
1996 April, seven of these spectra were centered on H$\alpha$, the
remaining three were taken at the blue end of the spectrum, to include
H$\beta$. We also obtained a H$\alpha$ spectrum from the AAT on 1996 March 2.
The SAAO spectra were all taken with the ITS spectrograph and RPCS
(Reticon) detector, with grating no.5, giving spectra covering the
region 6000-6800~\AA~ and 4400-5100~\AA~ at 0.5~\AA/pixel~dispersion. Data were acquired
in two channels (star and sky) simultaneously. The AAT spectrum was
taken with the 82cm RGO camera and TEK 1K CCD, with the 1200R grating,
giving a coverage of 6437-6677~\AA~at 0.234~\AA/pixel~dispersion. All
data were reduced using the Starlink supported {\sc figaro} package
(Shortridge 1991).\par
Table~2 shows a log of
observations. Six of the spectra (centred on H$\alpha$) are shown in
Figure~2. The H$\alpha$ profile is generally asymmetric, and
on one
occasion (1995 August) shows a clear double peak, with
V/R $>$1. There is a significant variation in the
equivalent width of the H$\alpha$ line. Throughout the period of observations, EW(H$\alpha$) has
decreased by $\sim$70 per cent, compared to a typical uncertainty in a
single value of $\sim$10 per cent. A decrease in the EW(H$\beta$) is
also clear. The results are further discussed
in Section 4.

%**************************OPTICAL PHOTOMETRY******************

\subsection{Optical photometry}

Str\"{o}mgren {\it uvby} photometry was taken from the catalogue of
the LTPV project (Sterken et al. 1995 and
references therein) covering the period 1982--1994. In addition, broad
{\it V} band photometry was taken from Pakull, Motch and Lub (1980,
hereafter PML1980) and from Hammerschlag-Hensberge et al. (1980). The $y$ data were transformed to standard $V$
magnitudes using the transform given by Sterken et al. (1995). The data
are plotted in Figure~3, along with
infrared {\it K} band magnitudes and EW(H$\alpha$) values. The {\it} K
band lightcurve consists of our data, plus data from Glass (1979) and
from Waters et al. (1988). H$\alpha$ values were taken from Cook and
Warwick (1987b) to supplement our own data. The data from
PML1980 show a rapid decline in optical luminosity during the first few months
of 1980. A further optical minimum is clearly seen in 1984 February 
(MJD~$\sim$46100), recovering to a maximum in 1990 April 
(MJD~$\sim$48000). The flux in all bands has been in decline from this
maximum until the date of the last observations.

\begin{figure}
\begin{center}{
 \epsfxsize \hsize
 \leavevmode
 \epsffile{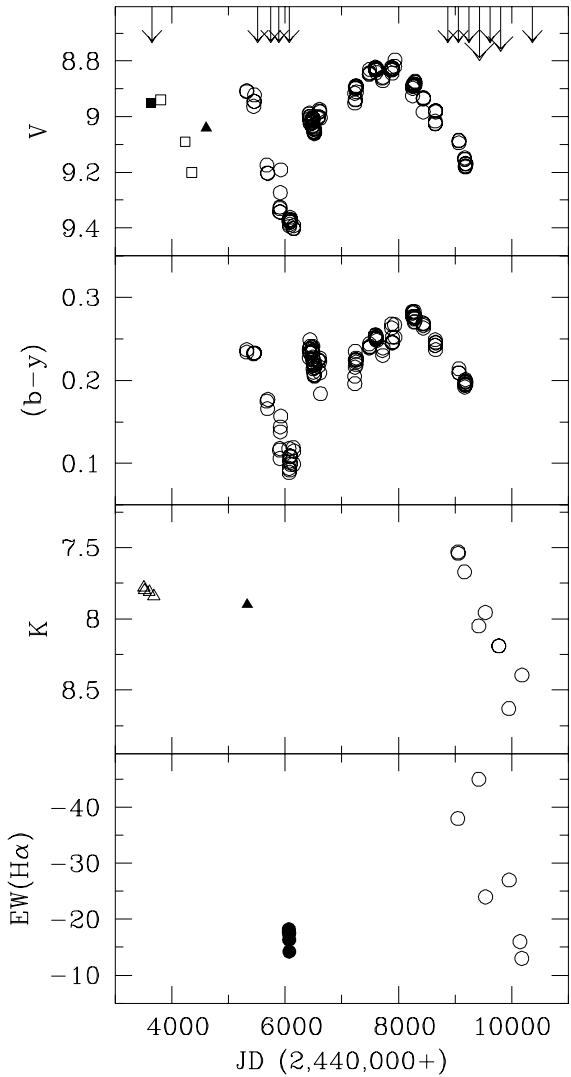}
}\end{center}
%\vspace{18.5cm}
\caption{{\bf Top and second panel:} Lightcurves of 4U1145-619 for the period 1978--1994, in
Str\"{o}mgren filter bands,  Open
squares represent data taken
from Pakull et al. (1980), open circles represent data from ESO's Long Term Photometric Variable
Project catalogues (Sterken et
al. 1995), closed triangles those of Densham \& Charles (1982), and closed squares represent data from
Hammerschlag-Hensberge et al. (1980).
{\bf Third panel:}~Infrared {\it K} band lightcurve from 1978--1996; the
open circles represent our own data, the open triangles those of
Glass (1979), and the closed triangles those of Waters et al. (1988).
{\bf Bottom panel:}~EW(H$\alpha$) plot, 1978--1996. Open circles again represent our own data,
closed circles those of Cook and Warwick (1987b). Arrows show the dates of detected
X-ray outbursts, and are sized to give a crude indication of the
intensity of the outburst.}
\label{strom}
\end{figure}

%***********************X-ray observations********************

\subsection{X-ray flux history}

An X-ray flux history has been compiled from the available
literature. All data were transformed to 2--10 keV fluxes with the
exception of the {\it EXOSAT} ME data which were given in this energy
range originally. The 2--10 keV fluxes were calculated assuming a
power law spectrum with parameters derived wherever possible from the
observations in question. This conversion is for comparison purposes
only, and flux values may be in error by as much as a factor of two
between one satellite and another. The 2--10 kev lightcurve is plotted in
Figure~4. There is a quiescent flux of $\sim$ 10$^{-10}$ ergs s$^{-1}$
cm$^{-2}$, and for the sake of discussion we shall regard an increase
by a factor of five in flux to be an outburst. The lightcurve then shows a
number of outbursts, recurring at the 186.5 day period. Two of the
outbursts stand out as the most luminous. The first, a 0.6-Crab (3--12
keV) outburst detected by the {\it Vela 5b} satellite in 1973 is
clearly an order of magnitude more intense than the remaining
outbursts, with the exception of a 0.5-Crab (20-40 keV) detection by the BATSE instrument on the {\it
CGRO} satellite, in March 1994. BATSE has detected six outbursts from the source
between MJD 48361 and 50231 (Scott M., private communication), though the one BATSE outburst plotted in
Figure~4 was much more intense than the remaining five
(Wilson et al. 1994; Scott
M., private communication). The most recent X-ray detection is a
0.1-Crab (2-12 keV) outburst detected by the {\it RXTE} satellite's
All Sky Monitor (ASM) between 1996 September 29 and October 8,
consistent with the 186.5 day period (Corbet \& Remillard
1996). The ASM data show a possible flare at the previous predicted
outburst epoch, but with far less significance.

%\begin{table*}
%\caption{X-ray fluxes for 4U1145-619 ******}
%\begin{tabular}{l c c c c c c c}
%\hline
%Date & MJD & Satellite & Energy &Phase & Flux & Derived 2--10 keV flux\\
%     &     &  &  keV && 10$^{-10}$ ergs s$^{-1}$ cm$^{-2}$ &10$^{-10}$ ergs s$^{-1}$ cm$^{-2}$ \\
%\hline
%1973 Apr 12 & 41785 & {\it Vela 5b} & 3--12 &0.00&  130 & 106\\   
%1979 Jul 13 & 44068 & {\it Einstein} & 0.2--2.0 &0.27& 0.086 & 1.26\\
%1992 Mar 6 & 48660 & {\it GRANAT} & 6--20 &0.89& $\leq$1.8 & $\leq$0.77\\
%1994 Mar 19 & & {\it CGRO (BATSE)} & 20--40 &0.00& 242 & 53.5 \\
%\hline
%\end{tabular}
%\end{table*}

\begin{figure*}
\begin{center}{
 \epsfxsize 1.0\hsize
 \leavevmode
 \epsffile{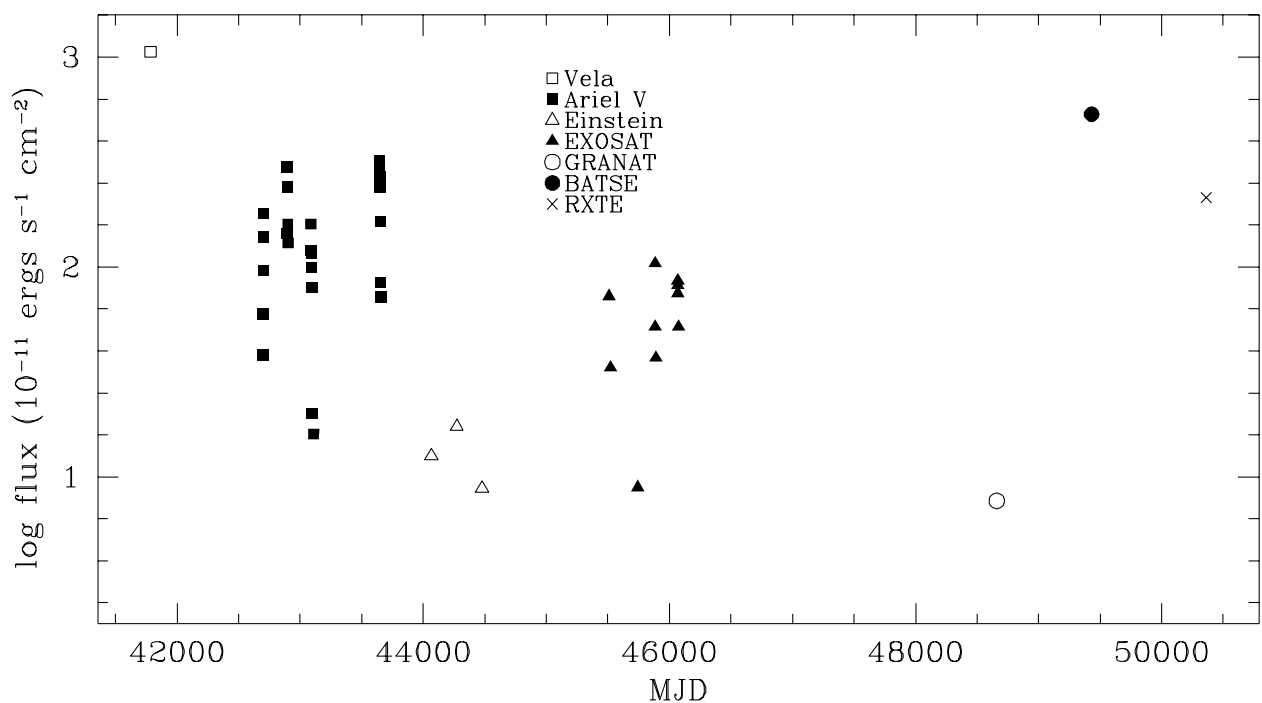}
}\end{center}
\caption{X-ray lightcurve of 4U1145-619 for the period 1973--1996, in
the 2--10 keV range. All values except the {\it EXOSAT} ME points are
calculated assuming a power law spectrum. This is to facilitate
comparison of measurements made in different energy bands, and values
may be in error by as much as a factor of 2 from one satellite to
another. Note that the GRANAT point is an upper limit. References for
X-ray data are: {\it Vela}, Priedhorsky \& Terrel 1983; {\it Ariel V},
Watson, Warwick \& Ricketts 1981; {\it Einstein}, Mereghetti et
al. 1987; {\it EXOSAT}, Mereghetti et al. 1987; Cook \& Warwick 1987a, 1987b;
{\it GRANAT}, Grebenev et al. 1992; BATSE, Wilson et al. 1994;
{\it RXTE}, Corbet \& Remillard 1996.}
\label{xray}
\end{figure*}

\section{Astrophysical parameters}

%*****************************PHYSICAL PARAMETERS***********

By design, the {\it uvby$\beta$} photometric system is most suitable for
determining the stellar parameters in early-type stars. Unfortunately, 
the Be star case is complicated by emission from the circumstellar disc, 
so that the observed indices become functions of both the stellar parameters
and of the disc parameters. Fabregat \& Reglero \shortcite{juan} determined 
transforms for the {\it uvby$\beta$} indices to correct for circumstellar 
effects. The transforms are based upon the EW(H$\alpha$) parameter, which has
been shown to correlate closely with circumstellar continuum and Balmer
discontinuity emission (Dachs et al. 1986, 1988; Kaiser 1989).\\

The photometric $uvby$ data used to derive the physical parameters
were obtained from the first catalogue of the LTPV project at ESO (Manfroid et al. 1991), and correspond to
three observations on 2nd and 5th January 1985. The errors are the mean value of the rms 
deviations of the differential measurements of those comparison stars
having at least six observations in one run (Manfroid et al. 1991).
We chose the photometric data as closely as possible to the optical 
minimum of 1984 (see Figure~3) since it is in this phase that the emission 
from the envelope is expected to be minimum. Consequently, the photometric magnitudes
present the lowest contamination from the circumstellar emission. In addition, 
contemporaneus measurements of the H$\alpha$ equivalent width were available
from Cook \& Warwick (1987b). The method and calibrations used to determine
the physical parameters are explained in Reig et al. (1996a). In the present
work, however, the value of the $\beta$ index used corresponds
to the mean for main-sequence stars, according to Equation 2 of Balona (1994). 
Mass was estimated by means of Balona's (1994) formula in terms
of luminosity and effective temperature, which is based on the
evolutionary models of Claret and Gim\'enez (1992). The error quoted for
the mass estimate is a formal error.
The results are presented in Table~\ref{param}.

%\begin{table*}
%\label{1145phot}
%\caption{Photometric results used in the determination of the
%astrophysical parameters of V801Cen, taken from the first
%Long-Term Photometric Variable catalogue (ESO project)}
%\begin{center}
%\begin{tabular}{|cccccccccc|}
%\hline
%Date &DJ &V &b-y &m$_1$ &c$_1$ &$\sigma_V$&$\sigma_{(b-y)}$
%&$\sigma_{m_1}$&$\sigma_{c_1}$\\
%\hline
%020185 &6068.784&9.373&0.118&-0.007&-0.051&0.005&0.005&0.008&0.007\\
%050185 &6071.806&9.375&0.089&0.032&-0.065 &0.005&0.005&0.008&0.007\\
%050185 &6071.807&9.370&0.094&0.027&-0.067 &0.005&0.005&0.008&0.007\\
%\hline
%\end{tabular}
%\end{center}
%\end{table*}

\begin{table}
\label{param}
\caption{Derived astrophysical parameters of V801 Cen, the optical counterpart 
to the Be/X-ray binary 4U 1145-619}
\begin{center}
\begin{tabular}{lc}
\hline
\hline
Spectral type 		&B1Ve 		 \\
E(B-V) 			&0.29$\pm$0.02 	\\
T$_{eff}$ 	  	&2.55$\pm$0.15 ($10^4$ K)	\\
Mass 		    	&13$\pm$2 (M$_{\odot}$)	\\
Radius 		 	&8$\pm$2 (R$_{\odot}$)	 \\
M$_V$			&-3.1$\pm$0.5 \\
M$_{bol}$		&-5.6$\pm$0.5\\
$\log g$ 		&3.8$\pm$0.2\\
Distance		&3.1$\pm$0.5 kpc\\
\hline
\end{tabular}
\end{center}
\end{table}

\section{Discussion}

%*****************************CIRCUMSTELLAR MATERIAL**********

\begin{figure}
\begin{center}{
 \epsfxsize 1.0\hsize
 \leavevmode
 \epsffile{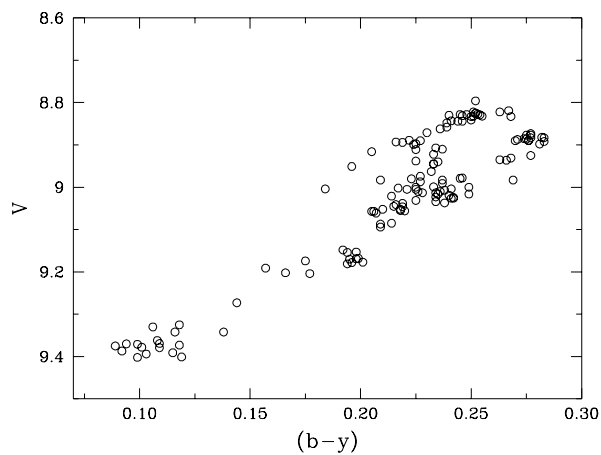}
}\end{center}
\caption{Colour-magnitude plot for V801 Cen, using $y$ and $(b-y)$
data taken from ESO's Long Term Photometry of Variables Project
catalogues (Sterken et al. 1995).}
\label{b-y}
\end{figure}

\subsection{Optical and infrared variations}
The infrared lightcurves in Figure~1 show the luminosity of the
source to be decreasing in all bands throughout the period of the
observations. Short term variations aside, the long term trend is a
decrease of $\sim$1 mag. The change in magnitude increases with
wavelength, with the change in {\it J,H,K} and {\it L} approximately
0.7, 0.9, 0.9 and 1.3 respectively. The H$\alpha$ and H$\beta$
emission has been in decay during the same period, the one data point in exception being that of
1994 March 7, which is just 10 days (0.05 phase) before the maximum of
the large X-ray outburst detected by BATSE. Observations have shown that the Balmer
emission and the infrared excess characteristic of Be stars originate
in the same circumstellar disc (Dachs and Wamsteker, 1982). Hence a decrease in disc size would result in the observed
variations in the infrared magnitudes and the H$\alpha$ and H$\beta$
emission. Also, as the disc's continuum is cooler than that of the
photosphere, it is at longer wavelengths that we expect to see the largest
variation in magnitude, again in agreement with the data.\par
In order to investigate the implications of the optical and
infrared data presented here in the context of the past behaviour of
the star, data were taken from the catalogue of ESO's Long Term
Photometry of Variables project (Sterken et al. 1995). The data
(Str\"{o}mgren $uvby$ photometry, covering the period 1982 to
1994) along with derived indices are shown in Figure~3, and were
described in Section 2.3. Of note is the scale of the long term
variability. The range of {\it V} magnitude is $\Delta${\it
V}$\sim$0.6, whilst $\Delta$({\it b--y})$\sim$0.2. As ({\it b--y}) increases
with optical luminosity, again the data indicate that the variations
are greater at longer wavelengths. Figure~5 shows explicitly
the relationship between luminosity and colour, represented by $V$ and
$(b-y)$ respectively. It is clear that in the high state, the
($b-y$) index shows the star to be redder than in the low state. As with the infrared photometry then, the size of the
variations and their dependance upon wavelength implies that they are the result of changes in
the size of the circumstellar disc. Hence, with the new infrared and
optical data we now have an indication of the way in which the disc
has decayed, recovered and decayed again over the past 13 years from 1982 December to
1996 April.\par
The complete data set shows three disc loss episodes, corresponding to
the two previously observed optical minima, and the current optical decay. Although the
infrared and spectroscopic data in the period before our own
observations is sparse, they are consistent with the timescale of disc
loss and recovery suggested by the optical lightcurve.
Similar
episodes of disc loss have been observed in other Be/X-ray binary
systems. Most notably, observations of X Persei have shown the
H$\alpha$ line changing from emission to absorption over a period of
months (Norton et al. 1991). In the case of 4U1145-619, complete disc
loss is clearly not the case at present, despite the decay in
strength, because both the H$\alpha$ and H$\beta$ lines remain in emission.

\subsection{X-ray behaviour}
In Figure~3 the epochs of X-ray outbursts are plotted as
arrows above the $V$ band lightcurve. Again we define an outburst as
an increase in flux by a factor of five above the quiescent flux of
$\sim$10$^{-10}$ ergs s$^{-1}$ cm$^{-2}$. Six BATSE detections are plotted, these corresponding to the six penultimate arrows (Scott M. private
communication). There appears to be a
correlation between the X-ray and optical behaviour, with X-ray
activity increased during periods of optical decline. Although we cannot
however rule out the possibility that the apparent correlation is an
artifact of the epochs of the X-ray observations, such a correlation
is not unexpected, as the X-ray emission is fueled by the material in
the varying disc. If the material lost from the disc is dissipated
away from the star, rather than falling back onto the surface, then
some fraction of this material should accrete onto the neutron star
providing additional fuel for X-ray emission. This scenario was
suggested by Roche et al. (1993) to explain similar correlations in
the X-ray and optical/infrared behaviour of the Be/X-ray binary X
Persei. Many Be/X-ray binary systems however show correlations of the
opposite nature, with increased X-ray activity coinciding with
optically bright phases (4U0115+63, Negueruela et al. 1997; A0538-66,
Corbet et al. 1985). A distinguishing factor between these two groups
is the significant difference in orbital period. Both 4U1145-619 and X
Persei are long period binaries, with wide orbits, 4U0115+63 and
A0538-66 each have periods less than 30 days, with much smaller
orbits. In the case of the long period binaries, the neutron star may
not become immersed in the disc during periastron passage, but rather
accrete from the material that is lost radially from this disc during
phases of decline in the Be star's activity. Such a scenario would be
consistent with spherical accretion.\par
In such a binary, where the neutron star does not accrete directly
from the Be star's disc, but from material lost radially from it, we
might expect any orbital modulation of X-ray flux to be of smaller
amplitude and smoother in profile than in
systems where the neutron star becomes immersed in the disc at
periastron. However in the case of 4U1145-619 the orbital modulation in the
X-ray lightcurve is significant, with outbursts lasting less than 0.1
phase and increases in flux of an order of magnitude above
quiescence. If this sharp modulation were caused by centrifugal
inhibition of accretion (Stella, White \& Rosner 1986), then we should
not detect quiescent flux from the source. Corbet (1996)
showed that X-ray emission may originate from the magnetosphere of the
neutron star, even
when accretion onto the neutron star surface was prohibited. We
note that Corbet finds in the case of 4U1145-619, that this emission
should be a factor of 7375 less than the minimum emission from
the neutron star's surface. The quiescent flux observed in 4U1145-619
is only a factor of $\sim$5 less than the normal outburst luminosity,
suggesting that accretion onto the neutron star's surface is still
occurring throughout the entire orbit. The sharp modulation may indicate an inclination between the planes of the
orbit and the Be star's disc, the sharp modulation occuring as the
neutron star's orbit crosses the plane of the disc at
periastron.

\subsection{Accretion mechanisms}
The source's relatively low
luminosity, wide orbit, and low pulsed fraction during `minor'
outbursts points towards spherical accretion as the source of energy
for X-ray emission from 4U1145-619 (White et al. 1983). Arons \& Lea (1980) suggested that the area of
the accretion hot spot in accreting neutron stars increases as the X-ray luminosity decreases. For low-luminosity
sources (L$_x$ $\leq$ 5 $\times$ 10$^{36}$ erg s$^{-1}$), as is the
case with 4U1145-619 during minor X-ray outbursts (see
Figure~4), assuming the derived distance of 3.1 kpc, the accretion area can be as large as the entire surface of the 
neutron star. Therefore, the angular momentum of the flow in 4U1145-619 
would be low and material would build up outside the
magnetosphere before penetrating and falling unevenly over most of
the surface of the neutron star.\par
The large outburst of 1994 March however allows an alternative scenario. At the maximum of the
major outburst, the total flux detected by BATSE was 0.5-Crab
(20--40 keV), with a
significant phase averaged pulsed flux of 0.3-Crab (Wilson et al. 1994), suggesting that in
this instance at least, accretion was concentrated to a greater degree onto the magnetic
poles. Such a difference in the pulsed fraction of the total flux
could be interpreted as evidence for the formation of a short-lived accretion disc, which would allow more
efficient binding of material to the magnetic field lines. In support
of this hypothesis, the spin period of the neutron star was seen to
decrease during the 1994 March outburst, whilst no significant spin-up
was detected during the five less intense outbursts
seen by BATSE between MJD~48361 and MJD~50231 (Scott M., private
communication). This
scenario would require a change in the circumstellar environment to
account for the lack of accretion disc formation at other periastron
passages. This requirement would be fulfilled by having a greater density of
circumstellar material resulting from the proposed mass ejection event from
the Be star. Further evidence for the formation of a temporary
accretion disc comes from the fact that the March 1994 outburst
continued at detectable levels longer than usual after maximum
flux. Mereghetti et al. (1987) however give an alternative
suggestion that such X-ray lightcurve profiles could be due to the compression of
accreting material in the bow shock of the neutron star, followed by
the accretion of more dilute ``downstream'' material. We also note
that shortly before the BATSE
detection in March 1994, the EW(H$\alpha$) strengthened to -45\AA, with
no correlated rise in the infrared luminosity. Either a delay exists
between the reaction of the disc's Balmer emision and infrared excess,
or an additional H$\alpha$ component was present, with a source
discrete from that of the infrared excess. An accretion disc
could produce this H$\alpha$ emission with no correlated change in the
infrared magnitudes.
\subsection{Astrophysical parameters}
Comparing the astrophysical parameters calculated in Section~3 to
previously published values yields few incompatibilities. The spectral
type of B1 Ve is in agreement with previous classifications based on spectroscopic results. Previous
values for {\it E(B--V)} have been in the range 0.25$\leq${\it
E(B--V)}$\leq$0.45, our value of 0.29 sits comfortably in the
middle. The temperature of the star has been determined from a model
atmosphere fit to an ultraviolet spectrum by Bianchi and Bernacca
(1980). Their result of 22,000 K is not consistent with our value of
25,500$\pm$1,500 K (Bianchi and Bernacca quote no errors), but we must
consider the difference of the methods used. The method used here has
removed circumstellar effects using a transform involving the
EW(H$\alpha$) index, whereas the method of Bianchi and Bernacca,
whilst not corrected for circumstellar effects, was applied in the
ultraviolet spectral region, where circumstellar effects may be
expected to be minimal. However, Kaiser (1989) notes that fitting
model atmospheres to Be star spectrophotometric measurements reveals
systematic differences between the model and observations, and
thus demonstrates the existence of radiation emitted by the envelope
at ultraviolet wavelengths. The continuum emission of the disc is
redder than that of the photosphere of the underlying Be star, and
hence unless circumstellar effects are taken into account, stellar
temperatures will be underestimated. The result which appears most
discrepant with previously published values is the distance of 3.1
kpc, a factor of two larger than the generally accepted 1.5 kpc to
this source (Hammerschlag-Hensberge et al. 1980; Lamb et
al. 1980). Again there are significant differences in the methods
employed in calculating the distance. Hammerschlag-Hensberge et
al. find 1.6 kpc from the dereddening necessary to fit observed
ultraviolet fluxes with expected spectra, but state that since colour
excess is known to be insensitive to distance at this particular
galactic longitude, the distance quoted may be in error by a factor
of three or more.\par

\section{Conclusions}
The Be/X-ray binary 4U1145-619 is exhibiting signs of disc loss in
optical and infrared wavelengths, after two previous such events since
1982. Whilst the infrared excess has decreased by $\Delta${\it
K}$\approx$1, the disc does not appear to have been lost completely,
as the H$\alpha$ line still shows emission, the last observations
showing EW(H$\alpha$) = -13 $\pm$ 1.3 \AA. X-ray luminosities and
pulse profiles during normal outbursts are typical of those expected
for spherically accreting systems, though there is a possibility of the
formation of a short-lived accretion disc during larger outbursts.\par
Physical parameters calculated for the optical counterpart V801 Cen
show discrepancies with previously published values, but the
discrepancies can usually be explained in terms of the circumstellar
effects which need to be corrected for in the case of Be
stars. Contemporaneous infrared and optical observations over the
course of the next optical rise, combined with X-ray observations
should confirm the existance or otherwise of any optical/X-ray
correlations, and lead to a more complete understanding of the
mechanisms fuelling this source, and long period Be/X-ray binaries in general.

\subsection*{Acknowledgments}

We are grateful to staff at the SAAO for their assistance, and to Tom Marsh and Chris Moran for obtaining the AAT
spectrum. Thanks also to the referee Reinhard Hanuschik for his comments on the original
manuscript. Much helpful information regarding BATSE
observations was provided by Matt Scott. This research has made use of the Simbad database
operated at CDS, Strasbourg, France. Part of the data reduction and analysis was carried out on the
Southampton University Starlink node which is funded by PPARC. JBS acknowledges the receipt of a research studentship from the
University of Southampton.

\bsp

\end{document}